# When Technology Outruns Science: A Scientometric Analysis of Large Language Model Evaluations

**Irene S. Gabashvili**

Aurametrix, USA

https://aurametrix.com

## Abstract

What happens when the technologies investigated by science evolve faster than science can communicate its findings? We examined 1,845 comparative large language model publication events from 2023 to 2026 to characterize the growing mismatch between technological progress and scientific dissemination. Median model age at publication more than doubled between 2023 and 2025, before declining modestly in 2026, while the upper tail continued to expand. Despite this apparent stabilization in calendar-time lag, technological succession became increasingly prevalent: by 2026, 97% of publications with identifiable model successors appeared after successor release. Among journal articles published after successor release with usable manuscript histories, 80.6% had already been overtaken by a newer model at receipt. Preprints featured substantially younger models even when compared with journal submissions rather than published articles (adjusted difference, 109 days; 95% CI, 66–152 days; $p < 0.001$). Neither assessor type nor evaluation complexity showed a statistically significant adjusted association with model age at receipt or preprint posting. These findings suggest that much of the technological aging of scientific evidence occurs upstream of journal editorial processing, limiting the extent to which faster peer review alone can address the problem. As technological development accelerates, the timeliness of scientific evidence may increasingly depend on when findings first become accessible rather than when they receive formal certification. Versioned research artifacts and earlier dissemination of evaluation results offer alternative approaches to scientific communication in rapidly evolving technological environments.

## 1. Introduction

Scientific knowledge is increasingly produced in an environment where the technologies under investigation evolve faster than the processes through which evidence is generated, validated, and communicated. This tension is particularly evident in the rapid succession of new general-purpose large language model (LLM) generations within a single research cycle. By the time a comparative evaluation reaches readers, the systems it examines may have already been superseded, potentially limiting its relevance to contemporary research and practice.

In an earlier systematic review of general-purpose LLM evaluations, we examined the breadth, methodological characteristics, and temporal evolution of published evidence (Gabashvili, 2025). The review highlighted the rapidly changing landscape of model evaluation and the challenges of interpreting evidence generated across different model generations. This raised a broader question extending beyond clinical applications: to what extent does the time required to produce and disseminate scientific research affect the technological currency of its findings?

Recent studies suggest that this temporal mismatch may be substantial. An analysis of 5,343 medical generative AI publications found that the newest model evaluated was a median of 14.2 months old at publication and that 94% of papers examined models for which newer releases were already available (Kim et al., 2026). Meanwhile, an analysis of scientific adoption across 62 LLMs and more than 108,000 citing papers found that successive model release years were associated with a 27% reduction in time to peak scientific adoption and a 23% reduction in scientific lifespan (Trišović, 2026). Together, these observations suggest that even stable publication delays may become increasingly consequential as technological lifecycles contract. The problem is therefore not simply how much calendar time elapses before research becomes available, but how much technological change occurs during that interval.

However, the technological age of published research is not determined by a single publication delay. Research may circulate as a preprint before journal submission, undergo peer review and revision, and appear in different publication formats at different times. The distinction between preprint availability, journal receipt, and final online publication is therefore important for understanding when scientific findings become accessible relative to the technological state of the field. Moreover, publication records often represent multiple dissemination events originating from the same investigation, each with a potentially different relationship to model release timelines.

To address these questions, the present study examines the temporal relationship between LLM development and scientific dissemination through a retrospective observational scientometric analysis of comparative LLM evaluations published between 2023 and 2026. We characterize the age of evaluated models at publication, determine whether technological successors were available at the time of dissemination, and investigate how technological age relates to different stages of the publication process, including preprint posting, journal receipt, and online publication.

### 1.1 Research aim

The primary aim of this study is to quantify technological obsolescence in scientific publications evaluating large language models and to characterize how the relationship between model development and scientific dissemination changes over time.

Additionally, we examined study characteristics that could potentially influence dissemination timelines, including research field, evaluation setting, primary system type, evaluation architecture and complexity, assessor type (human, automated, or hybrid), and literature identification pathway.

## 2. Methods

### 2.1 Study design and unit of analysis

We conducted a retrospective observational scientometric analysis of comparative large language model (LLM) evaluations to characterize the temporal relationship between model development and scientific dissemination. The primary outcomes were the age of the evaluated model at publication, the availability of a technological successor at publication, and, where manuscript histories were available, the distribution of technological age before and after journal receipt.

The individual publication event, rather than the underlying research study, was used as the unit of analysis. Distinct publications originating from the same investigation, including preprints, conference proceedings, and journal articles, were retained when they represented separate dissemination events. Duplicate bibliographic records referring to the same publication event were removed. This approach enabled examination of the technological timeliness of scientific communication across publication formats without requiring each observation to represent an independent research investigation.

### 2.2 Literature identification and corpus construction

Publication records were identified through two complementary approaches: review-derived identification and independent literature searches. Review-derived records were obtained from four existing systematic reviews, including published articles and preprints (Gabashvili, 2025; Rodrigues et al., 2026; Shool et al., 2025; Yang et al., 2026). These reviews provided an initial collection of primary studies evaluating LLMs, with substantial representation of clinical and healthcare applications.

Because the literature coverage of these reviews extended principally through early 2025, supplementary searches were conducted to extend temporal coverage and broaden disciplinary representation. Search strategies were informed by those employed in the source reviews but were not restricted to particular research disciplines or application domains. Independently identified records were obtained primarily through Dimensions database exports, supplemented by additional literature retrieval.

Records were classified according to their identification pathway:

- **Review-derived:** publications identified through existing systematic reviews, including published reviews and preprints.
- **Search-derived:** publications identified through independent literature searches.

The two identification approaches were used to broaden evidence retrieval while maintaining traceability of record provenance. The resulting corpus was intended to provide a heterogeneous, documented sample of comparative LLM research rather than an exhaustive census of all eligible publications. Differences in retrieval coverage across publication years were considered in the subsequent analyses.

## 2.3 Eligibility criteria and analytical cohort

The primary analysis focused on publications evaluating identifiable, general-purpose generative LLMs. Eligible evaluations included comparisons between multiple models or assessments of model performance against established benchmarks, reference standards, human performance, or other defined evaluation criteria.

Records were eligible when the evaluated model could be identified with sufficient specificity to assign a release date and when a corresponding publication event date could be established. Publications for which model versions were insufficiently reported were retained in the broader corpus but excluded from the primary temporal analysis.

Evaluations of encoder-only models, encoder–decoder architectures, and custom-developed systems were distinguished from evaluations of general-purpose autoregressive generative LLMs. These distinctions were introduced because the technological release histories and development trajectories of these systems differ substantially. Studies involving earlier language models, applied natural language processing, custom-model development, comparative reviews, and discovery-oriented methodologies were retained in secondary analytical strata where appropriate.

The initial corpus comprised 2,221 publication records. Of 1,962 records flagged as eligible, 1,864 were classified as comparative LLM studies. Screening selected-model identifiers and restricting publication events to 2023–2026 yielded 1,848 events. Restriction to journal articles, preprints, conference proceedings and book chapters retained 1,845 events. Deduplication by normalized DOI and online date, retaining the most recent selected-model release, yielded 1,845 publication events. Model ages were independently recomputed from dates; no negative publication ages remained. The latest observed publication event occurred on 28 November 2026; 2026 is a partial observation year.

For comparison with Kim et al. (2026), a health-related subset was defined using the original (unpooled) Analysis field: categories containing health, clinical, medical, dentistry, dermatology, oncology, radiology, surgery, neurology, pathology, or bioinformatics were retained. Records classified as Education were additionally included if the Source field contained clinical medicine, medical education, biomedical and clinical sciences, or health sciences (case-insensitive matching). This subject-based restriction was not a verification of PubMed indexing.

## 2.4 Bibliographic data extraction and verification

The extent and structure of available annotations differed across the source reviews. Records from Gabashvili (2025) contained comparatively extensive bibliographic, model-specific, and temporal annotations. For other reviews, supplementary bibliographic and methodological information was reconstructed from publication titles, DOIs, abstracts, available full texts, and publisher metadata.

Publication titles were matched to DOI identifiers where necessary. Abstracts and associated bibliographic metadata were retrieved programmatically using the Semantic Scholar Academic Graph API. Manuscript histories and publication dates were supplemented using the Europe PMC REST API, publisher websites, and available article metadata.

Extracted variables included publication title, DOI, retrieval source, publication format, research discipline, evaluation setting, evaluated model identifiers, selected-model release date, successor model and release date, and publication milestones. Additional methodological annotations were developed to characterize assessor type and evaluation complexity.

Model annotations supplied by the source reviews were not assumed to be complete or error-free. For example, among 761 records obtained from Shool et al. (2025), 728 (95.7%) contained model annotations. However, 210 of these records used an unspecified ChatGPT version as a baseline designation, leaving 518 records (68.1% of the original collection) with more specific model annotations before additional verification. Available article texts and supplementary materials were therefore examined to resolve ambiguous model identifiers. Inconsistencies in bibliographic records, duplicate entries, and disciplinary classifications were also corrected during corpus curation.

Data quality assurance involved approximately 15–16 successive review and verification passes combining manual examination, programmatic consistency checks, and LLM-assisted extraction and verification. LLM assistance was used to facilitate the identification of model versions, methodological characteristics, and potential metadata inconsistencies. Where accessible, ambiguous annotations were checked against the underlying publication or publisher metadata.

Full-text verification was not possible for all records because of subscription restrictions, automated retrieval barriers, and heterogeneous publication practices. Consequently, some annotations were based on abstracts, source-review information, or other accessible metadata. Unresolved model identifiers and methodological characteristics were not assumed to be known. Methodological classifications were treated as provisional annotations rather than independently validated measurements.

The curated dataset, data-processing scripts, metadata retrieval procedures, quality-control routines, and analytical code are documented in the repository associated with this study (https://github.com/Aurametrix/Scientometrics).

## 2.5 Selection of evaluated models and release dates

For publications evaluating multiple LLMs, the most recently released identifiable eligible model was selected as the reference model for temporal analysis. This approach was intended to characterize the most technologically recent component of each published evaluation rather than the age of its oldest model or the average age of all evaluated systems.

Model selection was based on the models documented as having been evaluated in the publication, rather than models mentioned only in its background or discussion. Release dates were assigned to the selected model or identifiable model version using the curated model-release information.

Where exact version identifiers could not be established but a sufficiently interpretable family-level designation and corresponding release date were available, records could be retained under the documented selected-model screening rule. Records for which no defensible release date could be assigned were excluded from the primary age analysis.

The resulting measure therefore represents the age of the newest identifiable eligible model documented in a publication. It does not necessarily represent the age of every model evaluated or establish that no more recent, insufficiently reported model was used.

## 2.6 Publication milestones and temporal outcomes

Publication timing was defined using the recorded first-online date of the corresponding publication event rather than the nominal journal issue date wherever these could be distinguished. For preprints, the relevant milestone was online posting; for journal articles, it was first-online publication. For conference proceedings, the recorded dissemination date could include a conference-date proxy when a more precise online publication date was unavailable.

Publication records carrying future-dated issue assignments were examined to distinguish nominal bibliographic dates from earlier documented online availability. Where early access could be established, the recorded availability date was used rather than the scheduled issue date. Remaining uncertainty concerning public accessibility and publication-date precision was acknowledged as a limitation of the metadata.

Selected-model age at publication was calculated as the difference in calendar days between the release date of the selected model and the recorded publication event date:

$$A_{pub} = t_{pub} - t_{release}$$

where $t_{pub}$ denotes the publication event date and $t_{release}$ denotes the selected-model release date.

Model age was interpreted as a measure of the temporal separation between technological release and scientific dissemination. It was not interpreted as a direct measure of research duration, experimentation time, manuscript preparation time, or publication-system inefficiency because the dates of model adoption and research initiation were generally unavailable.

## 2.7 Successor identification and technological succession

To distinguish elapsed calendar time from technological progression, the selected model was linked, where possible, to a subsequently released model within an identifiable developer or model lineage.

The primary successor analysis required a documented successor release date later than the selected-model release date and a relationship classified as a same-provider candidate, API-lineage replacement, same-family update, or same-line version candidate. Successor relationships were therefore operationally defined technological transitions rather than assertions that the selected model had been formally discontinued or functionally replaced for every application.

For each publication event with a qualifying successor, successor availability was coded as a binary outcome:

$$S(t) = 1, \text{ if } t > t_{successor}$$

$$S(t) = 0, \text{ if } t \leq t_{successor}$$

where t represents the publication milestone and t_successor denotes the successor release date.

A publication was classified as occurring after successor release only when its recorded date was strictly later than the successor release date. Events occurring on the same calendar date as successor release were classified together with events occurring before successor release. This conservative convention avoided inferring a temporal ordering within a calendar day when precise release and publication timestamps were unavailable.

Records without a resolved, dated successor relationship were excluded from the denominator of successor-based proportions rather than classified as having no available successor. A supplementary sensitivity analysis employed a more permissive date-based definition of successor eligibility.

Successor availability was interpreted as an indicator of technological succession and not as evidence that the evaluated model had become scientifically obsolete, that its successor performed better on the evaluated task, or that the original scientific findings had lost validity.

## 2.8 Decomposition of technological age using manuscript histories

For journal articles with identifiable manuscript receipt dates, model age at online publication was decomposed into two additive temporal intervals:

$$A_{receipt} = t_{receipt} - t_{release}$$

$$D_{post} = t_{pub} - t_{receipt}$$

such that

**A_pub = A_receipt + D_post**

Here, A_receipt represents the technological age accumulated by manuscript receipt, and D_post represents the subsequent receipt-to-publication interval.

The primary decomposition required chronologically consistent records satisfying:

**t_release ≤ t_receipt ≤ t_pub**

Records in which the selected model was released after manuscript receipt were examined separately because they could reflect the introduction of additional models during revision or inconsistencies in the reported publication history.

For articles with both usable manuscript histories and qualifying successor dates, successor availability was evaluated independently at manuscript receipt and online publication. Among articles published after successor release, the proportion for which the successor had already been released at manuscript receipt was calculated to distinguish technological transitions occurring before and after the reported submission milestone.

The release-to-receipt interval was not interpreted exclusively as research or manuscript preparation time, as it could include delayed model adoption, research design, data collection, experimentation, analysis, writing, and previous submission attempts. Similarly, the receipt-to-publication interval was not interpreted exclusively as peer-review duration because it could include editorial assessment, author revisions, production, and other publication activities.

Records in which the selected model was released after manuscript receipt were retained in the dataset because they may reflect technological updating during manuscript revision or discrepancies in manuscript-history metadata. These records were eligible for publication-age analyses but were excluded from the primary release-to-receipt decomposition because the selected model was not available at the initial submission milestone.

## 2.9 Characterization of the upper and lower tails

To examine exceptionally young and old-model evaluations while accounting for changes in publication timing across years and formats, publication events were classified relative to the empirical distribution of model age within each publication-year and publication-format stratum.

Events with model ages at or below the 25th percentile were assigned to the youngest group, whereas events with ages at or above the 75th percentile were assigned to the oldest group. Ties at the percentile thresholds were retained. Quartile-based classification was restricted to strata containing at least 20 publication events.

For journal articles with usable manuscript histories, release-to-receipt and receipt-to-publication intervals were compared between the youngest and oldest groups.

Differences in mean model age at publication were decomposed into the corresponding differences in mean model age at receipt and mean receipt-to-publication duration.

The proportion of the between-group difference already accumulated at manuscript receipt was calculated as:

$$\text{P_pre} = (\text{Mean_A_receipt_old} - \text{Mean_A_receipt_young}) / (\text{Mean_A_pub_old} - \text{Mean_A_pub_young})$$

where Mean_A_receipt_old and Mean_A_receipt_young represent mean model ages at manuscript receipt in the oldest and youngest groups, respectively, and Mean_A_pub_old and Mean_A_pub_young represent the corresponding mean model ages at publication.

The percentage of the between-group difference accumulated before manuscript receipt was obtained as:

$$\text{Percentage_pre} = 100 \times \text{P_pre}$$

Because the groups were defined using the observed publication-age outcome, this decomposition was descriptive and was not intended to estimate the causal contribution of editorial processing to publication delays.

## 2.10 Comparison of preprint posting and journal receipt

To investigate whether differences in model age between preprints and journal articles were already evident before formal journal processing, model age at preprint posting was compared with model age at manuscript receipt among journal articles with chronologically usable publication histories.

An exploratory ordinary least-squares regression estimated the difference in mean model age between the two dissemination milestones, adjusting for milestone calendar year and retrieval source. Milestone year was defined according to the receipt date for journal articles and the posting date for preprints.

The comparison involved distinct sets of publication events rather than systematically matched preprint and journal versions of the same manuscripts. Consequently, observed differences could reflect differences in research workflows, model selection, manuscript preparation, dissemination preferences, or other characteristics associated with publication format.

## 2.11 Classification of evaluation methodology

Additional methodological annotations were used to investigate whether human involvement or evaluation complexity was associated with technological age, particularly before journal editorial processing.

Assessor type was classified as automated, human, hybrid, or undetermined. Automated assessment referred to evaluations relying primarily on computational scoring, benchmark metrics, algorithmic procedures, or automated model-based assessment. Human assessment referred to evaluations relying primarily on judgments provided by human assessors, including domain experts. Hybrid assessment involved substantive contributions from both human and automated evaluation procedures.

Evaluation complexity was classified into three ordinal categories: basic, intermediate, and complex. The classification was intended to characterize the procedural and methodological demands of the evaluation rather than the intrinsic difficulty of the scientific question, the computational complexity of the evaluated model, or the actual labor invested in the study.

Complexity and assessor classifications were derived from available methodological descriptions, abstracts, full texts, and supplementary information. Where the available evidence did not permit a sufficiently defensible classification, the characteristic was coded as undetermined.

The principal methodological analysis used model age at manuscript receipt for journal articles and model age at posting for preprints. This combined outcome was intended to characterize technological age before the measured journal receipt-to-publication interval. Corresponding analyses at online publication were conducted for comparison.

## 2.12 Statistical analysis

Descriptive statistics included event counts, proportions, means, medians, interquartile ranges, sample variances, standard deviations, standard errors, and selected empirical quantiles. Conventional Student's t-based 95% confidence intervals were used where reported for descriptive means.

Because model-age distributions were asymmetric and included substantial upper tails, medians and empirical quantiles were emphasized in the characterization of temporal trends. Annual distributions were summarized using the median, 75th percentile, and 90th percentile, together with the proportion of selected models released at least 365 days before publication.

Unadjusted comparisons of model-age distributions between two groups were conducted using the Mann–Whitney U test. Comparisons involving multiple methodological categories were conducted using the Kruskal–Wallis test. These tests were interpreted as comparisons of distributions rather than exclusively as tests of median differences.

Exploratory multivariable ordinary least-squares regression was used to estimate adjusted differences in model age at online publication. The principal model included publication calendar year, publication format, retrieval source, research discipline, and evaluation setting as categorical predictors. The four source reviews were combined into a single review-derived category for regression analyses. Discipline and evaluation-setting categories represented by fewer than 25 events were pooled within their respective factors to reduce instability associated with sparse categories. Research discipline used the refined analysis field; sparse categories were pooled within the primary cohort.

Model fit was summarized using the coefficient of determination ($R^2$). The unique in-sample contribution of each predictor block was estimated as the reduction in $R^2$ following its removal from the full model:

$$\Delta R^2_j = R^2_{full} - R^2_{without_j}$$

where $R^2_{full}$ is the coefficient of determination for the complete model and $R^2_{without_j}$ is the coefficient of determination for the model excluding predictor block j.

Unique contributions were not assumed to be additive because predictor blocks could share explanatory information. Shapley decomposition was not recomputed in this revision.

For analyses of evaluation methodology, multivariable models incorporated publication year, publication format, retrieval source, research discipline, selected-model identity, assessor type, and evaluation complexity. Because model age was right-skewed, the outcome was log-transformed as follows:

$$Y = \ln(1 + A)$$

where A denotes model age in days at the relevant milestone and ln denotes the natural logarithm. Model and discipline categories with fewer than five observations were pooled within each methodological analysis to reduce sparse-category effects and improve model stability. The addition of one day permitted retention of observations with zero model age. Regression coefficients were exponentiated where appropriate and interpreted as proportional differences in the geometric mean of model age plus one day.

For a regression coefficient β, the corresponding percentage difference was calculated as:

$$\text{Percentage difference} = 100 \times [\exp(\beta) - 1]$$

Methodological analyses included the available provisionally coded observations, with undetermined classifications retained as separate categories where appropriate. Supplementary restricted analyses examined records with resolved classifications, including comparisons limited to explicitly human-assessed and automated evaluations.

Regression inference used HC3 heteroskedasticity-robust covariance estimates, two-sided t tests for individual contrasts, and robust omnibus Wald F tests for categorical predictor blocks. Approximate 95% confidence intervals were calculated using robust standard errors. No additional pairwise comparison families are reported in this revision.

The analyses were exploratory and were intended to characterize observational associations rather than identify causal determinants of technological aging. Nominal statistical significance was evaluated at the 0.05 level. Nonsignificant associations were not interpreted as establishing equivalence or demonstrating the absence of an effect.

## 2.13 Sensitivity analyses

Sensitivity analyses were used to examine the robustness of the observed temporal patterns to alternative cohort definitions, retrieval coverage, and analytical specifications. These included restrictions based on publication year, model-release year, and retrieval source; exclusion of the retracted publication; comparison of stored and independently recomputed model ages; upper-tail winsorization at the 99th percentile; and ln(1 + age) transformation.

Temporal trends were additionally examined among models released in 2023 or later to assess whether increasing publication age could be attributed primarily to the inclusion of earlier model generations. Analyses restricted to independently identified publications were used to examine whether changes in retrieval-source composition could account for the observed temporal patterns.

Successor-based analyses were repeated under a more permissive date-only eligibility definition. Secondary research strata were summarized separately because their inclusion criteria, technological time origins, and sampling conditions differed from those of the primary comparative LLM cohort.

To assess the influence of negative receipt ages, receipt-based analyses were repeated using model release (zero age), reported revision, acceptance, or combinations of these dates as alternative milestones. Subsequent intervals and milestone years were recalculated accordingly. These dates were not treated as verified dates of model incorporation.

## 2.14 Missing data and statistical dependence

Analytical sample sizes varied according to the availability of model release dates, qualifying successor relationships, manuscript histories, and methodological annotations. Successor-based proportions and manuscript-history analyses were calculated using records with the corresponding usable information. Undetermined methodological classifications were retained or excluded according to the requirements of the individual analysis.

Publication events were treated as independent observations in the primary statistical models. However, related publication versions and repeated evaluations of identical models could introduce dependence between observations. HC3 covariance estimation addressed heteroskedasticity but did not account for clustering by underlying study or evaluated model.

Accordingly, confidence intervals and statistical tests were interpreted as approximate, and the results were not assumed to represent population-level or causal estimates. Changes in retrieval coverage, model-family composition, and the opportunity for earlier models to accumulate age were considered when interpreting temporal associations.

## 3. Results

### 3.1 Corpus characteristics

The curated dataset comprised 2,221 records, including 1,864 eligible comparative LLM studies. The final primary cohort comprised 1,845 publication events: 1,622 journal article events (87.9%), 143 preprint events (7.8%), 77 conference proceeding events (4.2%), 3 book chapter events (0.2%). The latest recorded event occurred on 28 November 2026.

Median model age at online publication was 474 days (15.6 months; IQR, 284–721 days).

### 3.2 Temporal trends in model age at publication

Annual median model ages were 246 days in 2023, 482 days in 2024, 557 days in 2025, 531 days in 2026. Between 2025 and 2026 the 75th percentile changed from 819 to 1,048.2 days and the 90th percentile from 996 to 1,295 days.

The proportion of events evaluating models at least one year old was 2023: 15.8% (30/190); 2024: 76.1% (378/497); 2025: 66.4% (283/426); 2026: 68.7% (503/732).

Restriction to models released in 2023 or later yielded annual median ages of 201 days (2023), 428.5 days (2024), 532.5 days (2025), 523 days (2026).

**Table 1. Model age at online publication by year and publication format. Values are median days since model release, with event counts in parentheses.**

| Year | All events | Journal articles | Preprints | Conference proceedings |
|---|---|---|---|---|
| 2023 | 246 (190) | 261 (147) | 142 (35) | 266.5 (8) |
| 2024 | 482 (497) | 504 (445) | 226 (42) | 359 (10) |
| 2025 | 557 (426) | 564 (399) | 307 (17) | 368.5 (8) |
| 2026* | 531 (732) | 569 (631) | 228 (49) | 761 (51) |
| Overall | 474 (1,845) | 504.5 (1,622) | 218 (143) | 509 (77) |

*2026 through 28 November 2026. Book chapters are included in the overall totals but are not displayed separately.*

Retrieval coverage was different over time. 2023: 177/190 review-derived; 2024: 495/497 review-derived; 2025: 130/426 review-derived; 2026: 8/732 review-derived. Among search-derived publications, 2025 median 537, upper quartile 886.8 days; 2026 median 523, upper quartile 1,049 days.

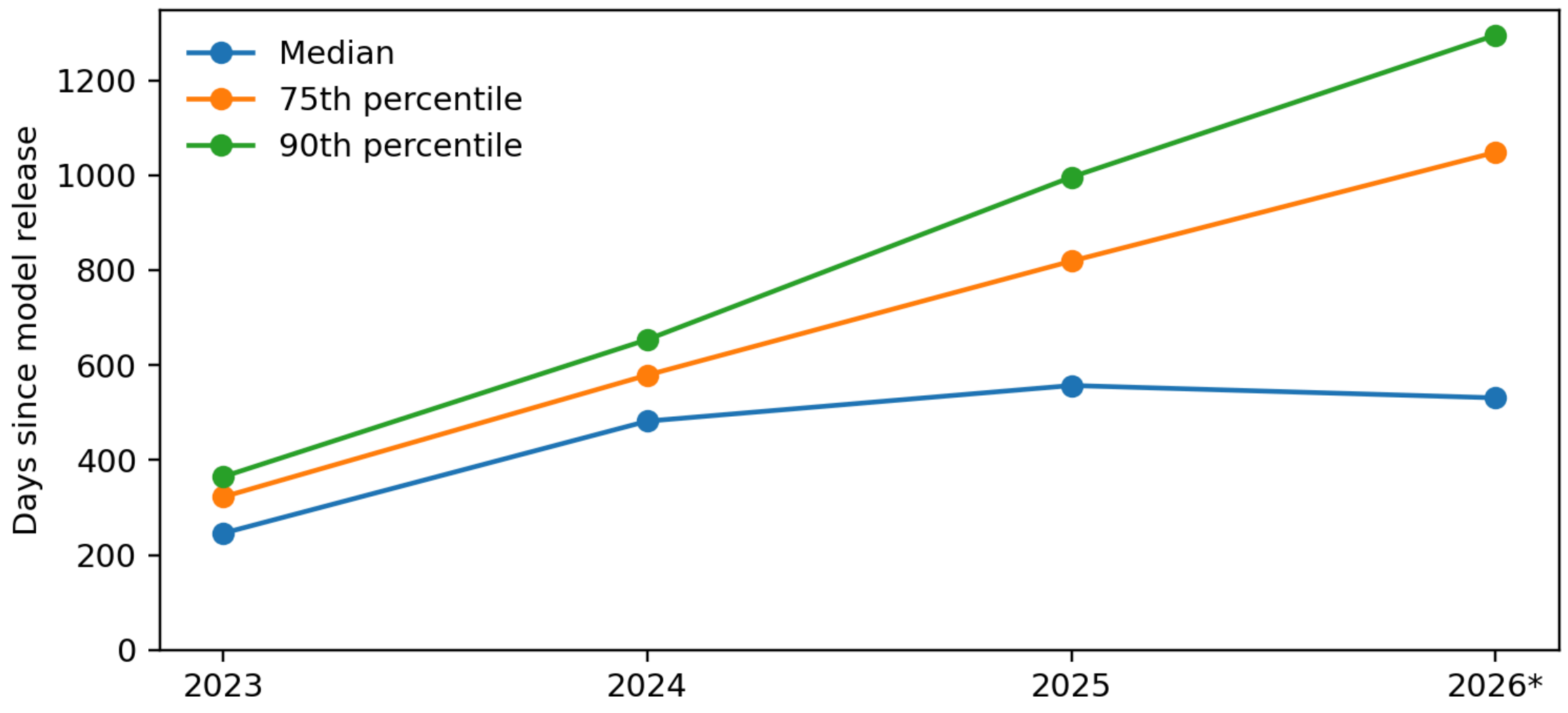

Figure 1. Model age at online appearance. Percentiles describe observed events. *Through 28 November 2026.

### 3.3 Successor availability at publication

A dated successor satisfying the prespecified lineage criteria was identifiable for 1,809 events. Of these, 1,648 (91.1%) appeared strictly after the successor release.

Annual proportions published after successor release were 2023: 68.9%, 2024: 90.4%, 2025: 91.9%, 2026: 97.0%.

**Table 2. Model age distribution and successor availability at online publication.**

| Year | Median age (days) | 75th percentile | 90th percentile | Published after successor release |
|---|---|---|---|---|
| 2023 | 246 | 322.5 | 365 | 131/190 (68.9%) |
| 2024 | 482 | 579 | 653.8 | 442/489 (90.4%) |
| 2025 | 557 | 819 | 996 | 385/419 (91.9%) |
| 2026* | 531 | 1,048.2 | 1,295 | 690/711 (97.0%) |

*Denominators for successor availability include only records with resolved, dated successor relationships. Publication on the successor release date was not classified as occurring after successor release.*

The increasing prevalence of successor availability, despite the stabilization of median model age in the most recent observation period, illustrates the distinction between elapsed calendar time and technological succession. The identification of a successor does not, however, establish that the earlier model was retired, that the successor performed better on the evaluated task, or that the original findings had become invalid.

### 3.4 Where technological age accumulates: before and after journal receipt

Of 1,622 journal articles, 556 (34.3%) had an identifiable receipt date. In 11 articles the selected model was released after receipt. Chronological and receipt-eligibility restrictions retained 545 articles for the primary decomposition.

Model releases after receipt may also reflect additions during revision or discrepancies in manuscript-history metadata; these records were excluded from the primary decomposition.

Median model age at receipt was 299 days and median receipt-to-publication duration was 137 days. In 422 articles (77.4%), the former interval exceeded the latter.

25 articles spent at least one year between receipt and online publication.

Among 538 usable journal histories with qualifying successors, 398 (74.0%) had passed successor release at receipt. Of 494 articles published after successor release, 398 (80.6%) had already reached that condition at receipt.

Thus, 80.6% of these articles had undergone technological succession before receipt.

### 3.5 Contribution of pre-submission and post-submission intervals to the upper tail

To determine where the exceptionally long intervals accumulated, journal articles were classified into the youngest and oldest quarters of model age within their respective publication years.

The youngest and oldest quarters included 135 and 116 usable histories, respectively. Median ages at receipt were 110 and 719 days; median subsequent intervals were 79 and 168.5 days.

**Table 3. Decomposition of model age among journal articles in the youngest and oldest within-year publication-age quartiles.**

| Measure | Youngest quarter | Oldest quarter |
|---|---|---|
| Articles | 135 | 116 |
| Median model age at receipt, days | 110 | 719 |
| Median receipt-to-publication interval, days | 79 | 168.5 |
| Mean model age at receipt, days | 125.6 | 731.7 |
| Mean model age at publication, days | 219.5 | 938.6 |

The mean publication-age gap was 719.2 days, of which 606.1 days (84.3%) had accumulated at receipt; the remaining 113.1 days reflected differences in receipt-to-publication duration.

Although these groups were defined by the outcome and do not provide a causal estimate of the contribution of editorial procedures, the decomposition demonstrates that the longest observed publication-age trajectories were distinguished predominantly by differences that preceded journal submission.

### 3.6 Preprints and journal articles: differences before peer review

Median publication ages were 218 days for preprints, 504.5 for journal articles and 509 for conference proceedings. The journal-conference Mann–Whitney comparison yielded p = 0.558.

To investigate whether the younger models observed in preprints could be explained solely by the absence of journal peer review and editorial processing, model age at preprint posting was compared with model age at journal receipt.

Median age at journal receipt was 299 days compared with 218 days at preprint posting.

**Table 4. Model age at journal receipt and preprint posting by milestone year.**

| Milestone year | Journal receipt, median days (n) | Preprint posting, median days (n) |
|---|---|---|
| 2023 | 179 (127) | 142 (35) |
| 2024 | 403 (140) | 226 (42) |
| 2025 | 412.5 (160) | 307 (17) |
| 2026* | 302 (117) | 228 (49) |
| Overall | 299 (545) | 218 (143) |

*The overall journal sample includes one additional receipt occurring in 2022.* Milestone year is determined by receipt or posting date, not journal publication date.

Adjusted for milestone year and retrieval source, preprint posting was associated with a mean age difference of -109.4 days relative to journal receipt (95% CI, -152.4 to -66.3; p = 7.71e-07).

The difference between preprints and journal publications was therefore evident even before the measured journal receipt-to-publication interval began. Because the comparison involved distinct publication events rather than matched versions of the same manuscripts, it could not distinguish differences in research initiation, model selection, manuscript preparation, previous submissions or publication preferences.

### 3.7 Factors associated with model age at online publication

The multivariable linear regression explained 24.9% of observed variance in online publication age (n = 1845).

Unique contributions and robust omnibus tests are reported in Table 5.

**Table 5. Associations with model age at online publication in the multivariable linear regression.**

| Predictor | Unique contribution to $R^2$ (percentage points) | Robust omnibus p-value |
|---|---|---|
| Publication year | 7.36 | 7.39e-94 |
| Publication format | 3.76 | 3.35e-26 |
| Research discipline | 4.45 | 8.91e-10 |
| Evaluation setting | 0.84 | 0.038 |
| Retrieval source | 0.66 | 8.9e-06 |

*Unique contributions were estimated by removing each predictor block from the full model. Inference used HC3 heteroskedasticity-robust covariance estimates. Contributions do not sum to the total explained variance because predictors share explanatory information.*

Unique contributions describe in-sample explanatory information, not causal effects.

## 3.8 Does the evaluation methodology explain model age before publication?

The expanded methodological annotations permitted a more direct examination of whether human assessment or evaluation complexity was associated with the technological age accumulated before formal journal processing.

The analytical sample comprised 688 events (545 journal articles and 143 preprints).

## 3.9 Human versus automated assessment

Assessor type was classified for 278 of 688 events (40.4%); 410 remained undetermined.

Human and automated evaluations had median ages of 314.5 and 261 days, respectively (Mann–Whitney p = 0.0853).

**Table 6. Model age at journal receipt or preprint posting according to assessor type.**

| Assessor type | Events | Mean age, days | Median age, days |
|---|---|---|---|
| Automated | 205 | 326 | 261 |
| Human | 60 | 387.2 | 314.5 |
| Hybrid | 13 | 342.9 | 247 |
| Undetermined | 410 | 360.4 | 281.5 |

Adjusted human versus automated assessment was associated with a 15.7% difference in geometric mean age plus one (95% CI, -20.6% to 68.6%; p = 0.448).

Restriction to explicitly human and automated evaluations yielded p = 0.446.

At online publication, the corresponding adjusted human-versus-automated contrast yielded p = 0.00911.

These exploratory estimates are subject to incomplete assessor annotations and do not establish causal effects.

## 3.10 Evaluation complexity

Evaluation complexity was classified for 607 of 688 events (88.2%); 81 remained undetermined.

Median ages by complexity were basic: 265.5 days, complex: 275 days, intermediate: 296.5 days, undetermined: 252 days.

**Table 7. Model age at journal receipt or preprint posting according to evaluation complexity.**

| Evaluation complexity | Events | Mean age, days | Median age, days |
|---|---|---|---|
| Basic | 176 | 332.4 | 265.5 |
| Complex | 113 | 359.4 | 275 |
| Intermediate | 318 | 371.5 | 296.5 |
| Undetermined | 81 | 309.4 | 252 |

Kruskal–Wallis p = 0.146 including undetermined classifications and 0.182 excluding them.

Adjusted complexity omnibus p = 0.837. Relative to basic evaluations, intermediate and complex evaluations yielded p = 0.902 and 0.727, respectively.

## 3.11 Multivariable associations with model age at receipt or preprint posting

The multivariable log-age model explained 46.9% of observed variance. Table 8 reports the adjusted omnibus tests.

**Table 8. Adjusted associations with model age at journal receipt or preprint posting.**

| Predictor | Robust omnibus p-value |
|---|---|
| Publication year | 2.94e-29 |
| Publication format | 0.724 |
| Retrieval source | 0.718 |
| Research discipline | 0.52 |
| Selected model | 1.96e-62 |
| Assessor type | 0.766 |
| Evaluation complexity | 0.837 |

Outcome: log(1 + model age in days). $R^2$ = 0.469; n = 688. P-values are nominal.

The strong associations with calendar year and model identity are partly inherent in the definition of model age, which depends on the release date of the selected model and the date of the publication milestone. These associations should not be interpreted as identifying causal determinants of research duration.

Methodological associations should be interpreted in light of missing annotations, model identity, calendar time and the exploratory design.

## 3.12 Comparison with secondary research strata

Secondary strata (all records; independently recomputed age where dates were available): Applied NLP: n = 15, median 1,402 days (dated n = 15); Comparative Review: n = 7, median 963 days (dated n = 7); Comparative Synthetic data generation by GPT-3.5 vs Clingen: n = 1, median 365 days (dated n = 1); Custom model: n = 1, median 1,296 days (dated n = 1); Custom, Encoder stratum: n = 9, median 0 days (dated n = 9); Discovery search stratum (Methodology): n = 1, median 524 days (dated n = 1); Discovery-search stratum: n = 12, median 245.5 days (dated n = 12); Encoder stratum: n = 51, median 1,366 days (dated n = 51); LLM Version not reported: n = 246, median 75 days (dated n = 1).

Secondary strata were not subjected to the primary selected-model screen.

These secondary strata were identified under different sampling conditions and, in some instances, used different technological time origins. Their distributions were therefore treated as descriptive rather than as directly comparable estimates of research or publication speed.

## 3.13 Sensitivity analyses

Release in 2023 or later: 2023=201, 2024=428.5, 2025=532.5, 2026=523; Search-derived: 2023=326, 2024=357.5, 2025=537, 2026=523; 2024–2026: 2024=482, 2025=557, 2026=531; Date-only successor definition: 1678/1841 (91.1%); Stored versus recomputed publication age: 0 mismatches; 99th-percentile winsorized regression $R^2$=0.254; log-age regression $R^2$=0.271. Retracted article already excluded by format restriction.

Of the 556 journal articles with identifiable receipt dates, 11 (2.0%) had negative model ages at receipt; five had reported revision dates and all 11 had acceptance dates. Across the five alternative-milestone analyses (550–556 journal articles), the proportion of the oldest–youngest mean publication-age gap attributable to model age at the analytical milestone ranged from 83.9% to 84.9% (primary analysis, 84.3%). Among articles published after successor release, the proportion already past successor release at the milestone ranged from 79.1% to 80.4% (primary analysis, 80.6%). The adjusted preprint age difference ranged from -107.1 to -103.6 days (all $p < 0.001$); adjusted assessor and complexity omnibus tests remained nonsignificant. The principal findings were robust to these alternative treatments.

The reconstructed screening and grouping rules are documented in the accompanying analysis script.

## 3.14 Summary of findings

The annual median and upper-tail trajectories differed: 2025 and 2026 median ages were 557 and 531 days, respectively; corresponding 90th percentiles were 996 and 1,295. In the journal-history subset, 84.3% of the oldest–youngest mean age gap was present at receipt. Preprint posting was associated with a 109.4-day lower mean model age than journal receipt after adjustment for milestone year and retrieval source ($p < 0.001$). Adjusted assessor and complexity omnibus p-values were 0.766 and 0.837.

The results locate much of the observed technological aging upstream of journal editorial processing, but the available dates do not resolve the relative contributions of delayed model adoption, research execution, manuscript preparation and previous submission attempts.

# 4. Discussion

## 4.1 Principal findings

The technological currency of published research continued to deteriorate even as median model age stabilized. By 2026, 97% of publications with identifiable successors appeared after successor release, while the upper tail of model age continued to expand. These findings complement evidence that successive LLM generations have progressively shorter scientific lifespans (Trišović, 2026).

Much of the observed technological aging preceded journal receipt. Among articles published after successor release with usable manuscript histories, 80.6% had already reached that condition at receipt, and 84.3% of the mean publication-age gap between the oldest and youngest groups had accumulated by this milestone. Preprints featured models approximately 109 days younger than those documented at journal receipt after adjustment for milestone year and retrieval source. However, publication format was not statistically significant in the more extensively adjusted log-age model, which also used publication year rather than milestone year. The association is therefore sensitive to analytical specification.

The preprint–journal difference raises questions about research practices before submission. Delayed model adoption, institutional approval procedures, coordination among collaborators, manuscript preparation, journal-specific formatting requirements, and previous submission attempts may all contribute. Some of this time may improve scientific rigor, whereas other activities may add little scientific value. A survey reported a median of 14 hours spent formatting each manuscript across the publication process (LeBlanc et al., 2019). Neither assessor type nor evaluation complexity showed a significant adjusted association with model age at the earlier dissemination milestone, although incomplete methodological annotations limit this interpretation.

## 4.2 Comparison with Prior work

An analysis of 5,343 medical publications naming a specific generative-AI model reported a median model age of 14.2 months at publication and newer releases available for 94% of papers (mean, 15.7 months, Kim et al., 2026).

The present comparative-evaluation cohort had a median age of 15.6 months and successor availability of 91.1% among lineage-evaluable events. Restriction to health-related publication events (n = 1,426) yielded a median of 14.2 months and a mean of 16.0 months, closely corresponding to the 14.2-month median and 15.7-month mean reported by Kim et al. (2026). Despite differences in sampling, eligibility criteria, and successor definitions, the two studies document a similar magnitude of technological aging in health-related research.

## 4.3 Implications for scientific communication

Accelerating manuscript production alone may not resolve the mismatch between technological development and scientific dissemination. Gartenberg et al. (2026)

documented increasing submission volumes and declining writing quality at Organization Science following the introduction of generative AI, illustrating how reduced manuscript-production costs may increase pressure on scientific evaluation.

An alternative is to separate the release of research evidence from its formal certification. Versioned research artifacts containing model identifiers, evaluation dates, protocols, code, outputs, and supporting evidence could be released before completion of the conventional publication process, where sharing is permitted. Emerging approaches, including Agent-Native Research Artifacts (Liu et al., 2026) and Paper2Agent (Miao et al., 2026), illustrate how research publications could become executable and interactive resources rather than exclusively narrative documents.

Such approaches need not replace human scientific judgment. Instead, reducing the effort devoted to manuscript preparation and routine verification could allow greater attention to originality, methodological validity, interpretation, and scientific significance. Technological currency should not become a performance metric that discourages replication or the deliberate evaluation of older models. These proposed publishing practices were not evaluated in the present study.

### 4.4 Limitations

This is a curated event sample. A model released late in 2022 had little opportunity to accumulate age in a 2023 publication. The early cohort is consistent with early adoption, but the shorter window alone constrains its upper tail. Publication year also tracks a shift from review retrieval to extended searches. Recorded event dates, developer lineages and model identities vary in precision. Related versions may contribute correlated observations. Missing receipt dates and selectively coded assessors limit the populations to which those estimates apply.

The oldest published events document an observable tail. The number of studies still unpublished, their eventual publication times, the duration of writing, and the effectiveness of peer review cannot be recovered from these dates. Estimating time from preprint to journal would require a preprint inception cohort, linkage to journal versions, a common search cutoff, and right-censoring of unresolved records. Work that never becomes public requires a different sampling frame.

## 5. Conclusion

The technological aging of scientific evidence is not solely a consequence of slow peer review. In comparative LLM research, much of the difference between younger- and older-model evaluations was already present at journal receipt, while preprints tended to document more recent models even relative to journal submissions. These findings motivate closer examination of research and publication practices before editorial processing.

As AI reduces the cost of producing manuscripts, verification, interpretation, and scientific attention may become increasingly important constraints. Earlier dissemination of versioned, evidence-bearing research artifacts, accompanied by concise human-readable explanations and interactive computational resources, offers a direction worth investigating. Reporting the date of the last evaluation would help distinguish deliberate model reuse from delays in communicating completed results. The objective should not be simply to publish faster, but to shorten the interval between producing reliable evidence and making it scientifically useful, without compromising rigor or human judgment.

## Acknowledgements

AI tools, primarily ChatGPT 6.0 and 6.2 (Astra and Sol) and Gemini 4, provided substantial assistance with data analysis. ChatGPT was also used to improve the readability, clarity, and flow of the manuscript. The author retains full responsibility for the interpretation of results and the final content.

## Data Availability Statement

The Python scripts, datasets, and all associated annotations are publicly available at the GitHub repository: https://github.com/Aurametrix/Scientometrics.